\documentclass{interact} 
 
\usepackage{graphicx,epstopdf}
\usepackage{subfigure}
 
\usepackage[numbers,sort&compress,merge]{natbib}
\bibpunct[, ]{[}{]}{,}{n}{,}{,}
 
\theoremstyle{plain}

\theoremstyle{definition}

\theoremstyle{remark}

\begin{document} 
 
\articletype{RESEARCH ARTICLE} 
 
\title{Excitation of H$_{2}^{+}$ with one-cycle laser pulses: Shaped post-laser-field electronic oscillations, generation of higher- and lower-order harmonics}
       \author{\name{Guennaddi K.~Paramonov\textsuperscript{a}, \thanks{CONTACT: G.K. Paramonov. Email: gk.parmon@gmail.com}  
       Oliver~K\"{u}hn\textsuperscript{b},
       and Andr$\acute{\rm e}$ D. Bandrauk\textsuperscript{c}} 
\affil{\textsuperscript{a}Institut f\"{u}r Chemie, Universit\"{a}t Potsdam,  
Karl-Liebknecht Strasse 24-25, 14476 Potsdam, Germany; 
\textsuperscript{b}Institut f\"{u}r Physik, Universit\"{a}t Rostock, Albert-Einstein-Strasse 23-24,  
D-18059 Rostock, Germany; 
\textsuperscript{c}Laboratorie de Chimie Th$\acute{\rm e}$orique, Facult$\acute{\rm e}$ des Sciences, 
Universit$\acute{\rm e}$ de Sherbrooke, Sherbrooke, Qu$\acute{\rm e}$bec, J1K 2R1, Canada}} 
 
\maketitle 
 
\begin{abstract} 
Non Born-Oppenheimer quantum dynamics of H$_{2}^{+}$ excited by shaped one-cycle laser pulses linearly polarized along the molecular axis have been studied by the numerical solution of the time-dependent Schr\"{o}dinger equation  
within a  
three-dimensional model, including the internuclear separation, $R$, and the electron coordinates 
$z$ and $\rho$. Laser carrier frequencies corresponding to the wavelengths $\lambda_{l}=25$~nm through 
$\lambda_{l}=400$~nm were used and the amplitudes of the pulses were chosen such that the energy of H$_{2}^{+}$  
was close to its dissociation threshold at the end of any laser pulse applied. 
It is shown that there exists a characteristic oscillation frequency $\omega_{\rm osc} \simeq 0.2265$~au  
(corresponding to the period of $\tau_{\rm osc} \simeq 0.671$~fs and the wavelength of $\lambda_{\rm osc} \simeq 200$~nm) that manifests itself as a ``carrier'' frequency of temporally shaped oscillations of the time-dependent expectation values 
$\langle z \rangle$ and $\langle \partial V/\partial z \rangle$ that emerge at the ends of the laser pulses and exist   on a timescale of at least 50~fs. Time-dependent expectation values $\langle \rho \rangle$ and 
$\langle \partial V/\partial \rho \rangle$ of the optically-passive degree of freedom,  $\rho$, demonstrate post-laser-field  oscillations at two basic frequencies  
$\omega^{\rho}_{1} \approx \omega_{\rm osc}$ and $\omega^{\rho}_{2} \approx 2\omega_{\rm osc}$. 
Power spectra associated with the electronic motion show higher- and lower-order harmonics with respect to the driving field. 
\end{abstract}


\section{Introduction} 
 
 Coherent oscillations of bound particles (electrons or muons) that persist after  laser field excitation of a 
  molecule~ 
 \cite{Bandrauk:jpcA.2012.post-field,%
 BandrParamon:ijmpE.2014.muon,%
 ParKuehnBandr:jpcA.2016.H2p.post-field} 
 is an interesting new phenomenon, which  generalizes the well-known recollision model of Corkum 
 \cite{Corkum:93prl.Recollision} 
 to the case when high-order harmonics are generated due to coherent oscillations of bound electrons or muons after the end of the laser field that induced their motion. Coherent post-laser-field oscillations of bound  particles were found to exist both in ordinary (electronic) and muonic molecules, as shown  by numerical propagation of the respective non-Born-Oppenheimer time-dependent 
 Schr\"{o}dinger equations 
 \cite{Bandrauk:jpcA.2012.post-field,%
 BandrParamon:ijmpE.2014.muon,%
 ParKuehnBandr:jpcA.2016.H2p.post-field}. 
 
 Coherent post-laser-field electronic oscillations were first found in the heavy molecular ion T$_{2}^{+}$ 
 after its excitation by a UV laser pulse 
 \cite{Bandrauk:jpcA.2012.post-field}. 
 Subsequently 
 \cite{BandrParamon:ijmpE.2014.muon}, 
 coherent and shaped post-laser-field oscillations of a muon were found in muonic $dd\mu$ and $dt\mu$ molecules 
 after their excitation by super-intense attosecond soft X-ray laser pulses at the wavelength of $\lambda_{l}=5$~nm. 
 It was found, in particular, that only odd harmonics were generated in the homonuclear $dd\mu$ molecule  
 by the optically active $z$ degree of freedom, as suggested by the concept of inversion symmetry 
 \cite{Gross:prl.2001.HHG}. Only even harmonics were generated in $dd\mu$ by the optically passive, transversal $\rho$ degree of freedom, 
 which is excited only due to the wave properties of an electron. 
 In contrast, both odd and even harmonics were generated in the heteronuclear $dt\mu$ molecule  
 by the optically active $z$ degree of freedom due to inversion symmetry breaking 
 \cite{Gross:prl.2001.HHG}, 
 and both even and odd harmonics were generated in $dt\mu$ by the optically passive $\rho$ degree of freedom. 
 It was also shown in Ref.~\cite{BandrParamon:ijmpE.2014.muon} 
 that the appearance of coherent muonic oscillations in $dd\mu$ and $dt\mu$ after the end of the laser field 
 is a purely non-Born-Oppenheimer effect: the post-laser-field muonic oscillations did not occur if the Born-Oppenheimer approximation was employed. 
 
 In our recent work 
 \cite{ParKuehnBandr:jpcA.2016.H2p.post-field}, 
 shaped post-laser-field electronic oscillations were also found to exist in H$_{2}^{+}$  
 excited by two-cycle laser pulses at the wavelengths $\lambda_{l}=800$ and 200~nm. 
 It was shown, in particular, that there exists a characteristic oscillation frequency $\omega_{\rm osc}=0.2278$~au  
 (corresponding to the wavelength of $\lambda_{l}=200$~nm) that plays the role of a ``carrier'' frequency of  
 temporally shaped post-laser-field electronic oscillations both at $\lambda_{l}=800$~nm and at $\lambda_{l}=200$~nm. 
 In the present work, we investigate the non-Born-Oppenheimer quantum dynamics of H$_{2}^{+}$ 
 excited by shaped \emph{one-cycle} laser pulses, i.e. the shortest periodic excitation of H$_{2}^{+}$ is studied. 
 The laser carrier frequencies corresponding to the wavelengths $\lambda_{l}=25$, 50, 100 150, 200, 300 and 400~nm  
 are used   such as as to cover the domains of both $\omega_{l} \leq \omega_{\rm osc}$ 
 and $\omega_{l} > \omega_{\rm osc}$. 
 Similarly to our previous work  
 \cite{ParKuehnBandr:jpcA.2016.H2p.post-field}, 
 the amplitudes of one-cycle laser pulses are chosen such that the energy of H$_{2}^{+}$ after the ends  
 of the pulses are $\langle E \rangle \simeq -0.515$~au,  
 i.e., slightly below its dissociation threshold $E_{\rm D}=-0.5$~au. 
 
\section{Model, equations of motion, and numerical methods} 
 
 The three-body three-dimensional model with the Coulombic interactions 
 representing the ${\rm H}_{2}^{+}$ 
 excited by the laser field linearly polarized 
 along the $z$ axis 
 is shown in Figure~1. 
 The nuclear motion is assumed to be restricted to the 
 polarization direction of the laser electric field.  
 The electron (${\rm e}$) moves in three dimensions with conservation 
 of cylindrical symmetry. Therefore, only two electronic 
 coordinates, $z$ and $\rho$, measured with respect 
 to the center of mass of the two protons (${\rm p}$) 
 should be treated explicitly together with the internuclear 
 separation $R$. 
 
\begin{figure}[t] 
\begin{center} 
\includegraphics*[width=0.57\textwidth]{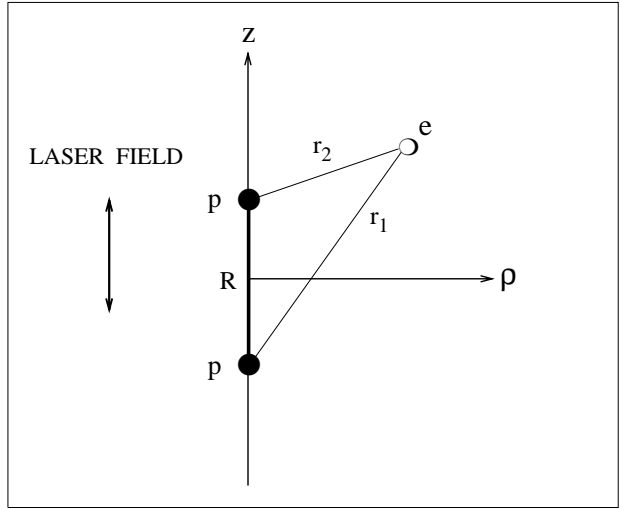} 
\end{center} 
\caption{The three-body three-dimensional model of ${\rm H}_{2}^{+}$ 
 excited by a laser field linearly polarized along the $z$ axis. 
 The internuclear distance is $R$, 
 the distances between the electron and 
 each of the two protons are $r_{1}$ and $r_{2}$, 
 see Equations~(\ref{rH}) and (\ref{rD}).} 
\end{figure} 
 
 The component of the dipole moment of ${\rm H}_{2}^{+}$ 
 along the $z$ axis reads 
\cite{Carrington:89jphB.HD} 
\begin{equation} 
 d_{z}(z) = - e z[1+M_{\rm e}/(2M_{\rm p}+M_{\rm e})], 
\label{DzH2} 
\end{equation} 
 where $-e$ is the electron charge, 
 $M_{\rm p}$ and $M_{\rm e}$ are the proton 
 and the electron masses, respectively. 
 The homonuclear ${\rm H}_{2}^{+}$ molecular ion does not have 
 a permanent dipole moment, therefore 
 its vibrational motion is excited only indirectly 
 due to electronic motion induced by the laser field 
 along the $z$ axis 
 \cite{KawataKonoFujimura:JCP1999.H2p.rho,%
 Paramon:05cpl.HH-HD,%
 Paramon:07cp.HH-HD-Muon}. 
 Electronic motion along the transversal $\rho$ axis occurs only due to 
 the wave properties of the electron. 
 
 The time-dependent non-Born-Oppenheimer Schr\"{o}dinger equation that governs 
 the quantum dynamics of ${\rm H}_{2}^{+}$ 
 in the classical laser field ${\cal E}(t)$ reads 
 \begin{displaymath} 
 i\hbar \frac{\partial }{\partial t} 
  \Psi   
  - \frac{\hbar^{2}}{2 m_{\rm n}} 
  \frac{\partial^{2} \Psi }{\partial R^{2}} 
  - \frac{\hbar^{2}}{2 m_{\rm e}}\biggl( 
   \frac{\partial^{2} \Psi }{\partial \rho^{2}}  
     + \frac{1}{\rho } 
 \frac{\partial \Psi }{\partial \rho }\biggr) 
 \end{displaymath} 
 \begin{equation} 
 - \frac{\hbar^{2}}{2 m_{\rm e}} 
 \frac{\partial^{2} \Psi }{\partial z^{2}} 
  +V(R,\rho,z)\Psi 
  - d_{z}(z){\cal E}(t)\Psi. 
 \label{TDSE1} 
 \end{equation}  
 In Equation~(\ref{TDSE1}), 
 $m_{\rm n}=M_{\rm p}/2$ 
 is the nuclear reduced mass, 
 $m_{\rm e}=2M_{\rm e}M_{\rm p}/(M_{\rm e}+2M_{\rm p})$ 
 is the electron reduced mass. 
 In the atomic units (au) used throughout the paper, we have: 
 $e=\hbar=M_{\rm e}=1$ and $m_{\rm e}\simeq 1$ and for the field 
 amplitude and intensity ${\mathcal E}_0=5\times10^9$~V/cm and 
 $I_0=3.5\times10^{16}$~W/cm$^2$, respectively.  
      
 The Coulomb potential reads 
\begin{equation} 
 V(R,\rho,z)=e^{2}(1/R - 1/r_{1} - 1/r_{2}), 
 \label{Coul} 
 \end{equation} 
where the electron-proton distances 
(see Figure~1) are 
\begin{equation} 
 r_{1}(R,\rho,z)=[\rho^{2} + (z + R/2)^{2}]^{1/2} 
 \label{rH} 
 \end{equation} 
 and  
 \begin{equation} 
 r_{2}(R,\rho,z)=[\rho^{2} + (z - R/2)^{2}]^{1/2}. 
 \label{rD} 
 \end{equation} 
 
 The time-dependent laser electric field 
 ${\cal E}(t)$ 
 is chosen as follows: 
 \begin{equation} 
 {\cal E}(t) = {\cal E}_{0}\sin^{2}(\pi t/t_{p})\sin (\omega_{l} t), 
 \, \, \, \, \, \, 0 \leq t \leq t_{p}, 
 \label{Lfield} 
 \end{equation} 
 where ${\cal E}_{0}$ is the amplitude, $t_{p}$ is the pulse duration at the base, 
 and $\omega_{l}$ is the laser carrier frequency. 
 With the symmetric (the $\sin^{2}$-type) envelope and the integer number of optical cycles at the base 
 (one optical cycle in the present case), 
 the laser electric field 
 ${\cal E}(t)$ 
 has a vanishing direct-current component, 
 $\int_{0}^{t_{p}}{\cal E}(t)dt = 0$, 
 and satisfies therefore  Maxwell's equations in the propagation region 
 \cite{ParamonSaalfrank:PRA2009.time-evol}. 
 Note before proceeding, that at a small number of optical cycles per pulse duration, 
 as in the present work for example, the carrier-envelope phase (CEP) may play 
 a very important role 
 \cite{Bandrauk:2002.PRA.HatomNc15}. 
 We shall address this issue in our future work. Here, we just assume 
 that the carrier-envelope phase is equal to zero, see Equation~(\ref{Lfield}). 
 
 The numerical methods used to solve the three-dimensional Equation~(\ref{TDSE1}) 
 have been described in our previous works 
 \cite{Paramon:05cpl.HH-HD,%
 Paramon:07cp.HH-HD-Muon}. 
 In particular, the dissociation probability has been calculated with 
 the time- and space-integrated outgoing flux for the nuclear coordinate $R$; 
 the ionization probabilities have been calculated 
 with the respective fluxes separately for the positive 
 and the negative direction of the $z$ axis 
 as well as for the outer end of the $\rho $ axis. 
 
 The size of the $z$-grid has been chosen such as to be substantially larger 
 than the maximum electron excursion along the $z$ axis, 
 $\alpha_{z} = {\cal E}_{0}/\omega_{l}^{2}$, 
 and the size of the $\rho$-grid has been chosen accordingly. 
 The maximum electron excursion, $\alpha_{z} =9.25$ au, corresponds to the field parameters used  
 at $\lambda_{l}=400$~nm (${\cal E}_{0}=0.12$~au and $\omega_{l}=0.1139$~au) 
 and the choice of the $z$ and $\rho$ grids has been based on this value. 
 Specifically, the three-dimensional wave-function $\Psi(t)$ of Equation~(\ref{TDSE1})  
 was damped with the imaginary smooth optical potentials, adapted from 
 \cite{Rabitz:94.jcp.HF}, 
 at $z<-263$~au, at $z>263$~au  
 and at $\rho > 239$~au for the electronic motion, 
 and at $R>23$~au for the nuclear motion. 
 
 Initially, at $t=0$, the ${\rm H}_{2}^{+}$ 
 was assumed to be in its ground vibrational and 
 ground electronic state. 
 The wave function of the initial state 
 was been obtained by the numerical propagation 
 of Equation~(\ref{TDSE1}) 
 in the imaginary time without the laser field (${\cal E}_{0}= 0$). 
 
 \section{Results}
 \subsection{Resonant properties, laser-driven dynamics, and free evolution of ${\rm H}_{2}^{+}$ on a short timescale} 
 
 As it was already mentioned, the amplitudes ${\cal E}_{0}$ of the one-cycle laser pulses used in the present work 
 have been chosen such that 
 the energies of ${\rm H}_{2}^{+}$ at the ends of the laser pulses were 
 slightly below the dissociation threshold ($E_{D}=-0.5$~au) 
 and similar, $\langle E(t=t_{p}) \rangle \approx -0.515$~au. 
 The amplitudes of one-cycle laser pulses required to achieve 
 the aforementioned energy are plotted in Figure~2 versus their wavelength $\lambda_{l}$ by curve~1. 
 For the sake of comparison, similar results obtained with two-cycle laser pulses are presented  
 in Figure~2 by curve~2. 
 
\begin{figure}[h] 
\begin{center} 
\includegraphics*[width=0.57\textwidth]{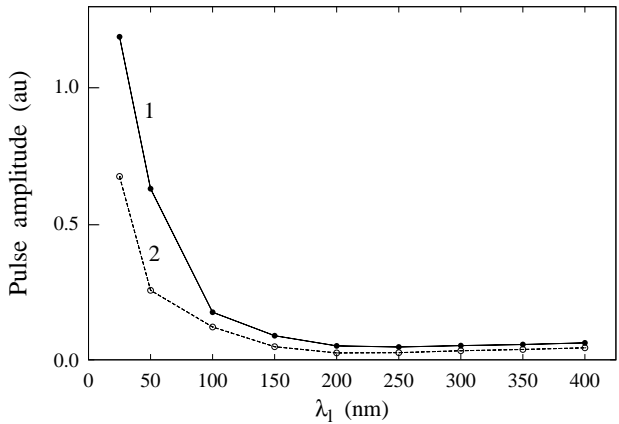} 
\end{center} 
 \caption{The amplitudes of one-cycle (curve~1) and two-cycle (curve~2)  
 laser pulses required to excite ${\rm H}_{2}^{+}$ from its ground vibrational 
 state $v=0$ to the energy of $\langle E \rangle \approx -0.515$~au.} 
\end{figure} 
 
 Two different domains of the laser wavelength $\lambda_{l}$ can be clearly distinguished in Figure~2. 
 The domain of a large change of the laser pulse amplitude ${\cal E}_{0}$ required to excite ${\rm H}_{2}^{+}$ 
 to the energy of $\langle E \rangle \approx -0.515$~au with the laser wavelength $\lambda_{l}$ 
 corresponds to $\lambda_{l} < 200$~nm (or $\omega_{l} > \omega_{\rm osc}$). 
 The domain of a relatively small change of ${\cal E}_{0}$ with $\lambda_{l}$ corresponds to $\lambda_{l} \geq 200$~nm 
 ($\omega_{l} \leq \omega_{\rm osc}$). 
 It is also seen from Figure~2 that the most efficient excitation of ${\rm H}_{2}^{+}$ 
 by one-cycle and two-cycle laser pulses, which requires the minimum laser electric-field amplitude, takes place at $\lambda_{l}^{\rm opt} \simeq 200$~nm, 
 which plays the role of a resonant, or the optimal laser wavelength. 
 It can be expected therefore that power spectra of electron oscillations resulting from excitation of ${\rm H}_{2}^{+}$ to  
 $\langle E \rangle \approx -0.515$~au at various laser wavelengths $\lambda_{l} \neq 200$~nm should  
 always contain harmonics corresponding to $\lambda \approx 200$~nm. 
 
 Indeed, it was shown in our recent work 
 \cite{ParKuehnBandr:jpcA.2016.H2p.post-field}, 
 where excitation of ${\rm H}_{2}^{+}$ with two-cycle laser pulses at $\lambda_{l} = 200$ and 800~nm was studied, 
 that the respective power spectra contain the strongest harmonics corresponding to $\lambda = 200$~nm,  
 the fourth harmonic at $\lambda_{l} = 800$~nm and the first (identical) harmonic at $\lambda_{l} = 200$~nm. 
 It will be interesting to check therefore whether the excitation of ${\rm H}_{2}^{+}$ with laser wavelengths 
 $\lambda_{l} < 200$~nm would result in the generation of lower-order harmonics. 
 
 An important problem of the laser-driven dynamics with few-cycle laser pulses 
 is the electron-field following. 
 According to the well-known recollision model of Corkum 
 \cite{Corkum:93prl.Recollision}, 
 electron follows the field out-of-phase: the expectation value  
 $\langle z(t) \rangle$ decreases when electric-field strength ${\cal E}(t)$ increases. 
 Previously, a perfect electron-field out-of-phase following on the level of expectation values 
 $\langle z \rangle$ 
 of the laser-driven electronic degree of freedom 
 have been found only in the infrared  
 \cite{Paramon:05cpl.HH-HD,%
 Paramon:07cp.HH-HD-Muon} 
 and near-infrared 
 \cite{Kono:04cp.H2p.nearIR} 
 domains of the laser carrier frequency, with a large number of optical cycles  
 per pulse duration $t_{p}$ being involved. 
 It was also found in our previous work 
 \cite{ParamonKuhBand:2011pra.H-H} 
 that electrons of the extended H-H system follow the applied laser field out-of phase 
 at the laser carrier frequency $\omega_{l} < 0.1$~au, and in-phase at $\omega_{l}=1$~au, 
 with the number of optical cycles per pulse duration being about 33.  
 
 The laser-driven dynamics of ${\rm H}_{2}^{+}$ excited by two-cycle laser pulses 
 was studied in our recent work 
 \cite{ParKuehnBandr:jpcA.2016.H2p.post-field}. 
 It was found, in particular, that at $\lambda_{l} = 800$~nm,  
 the expectation value $\langle z \rangle$ follows the field out-of-phase only approximately 
 (during the first optical cycle),  
 while at $\lambda_{l} = 200$~nm, the out-of-phase electron-field following 
 does not take place even approximately. 
 It would be moreover difficult to expect the out-of-phase electron-field following 
 for expectation values $\langle z \rangle$ at one-cycle excitation. 
 
 The laser-driven dynamics of ${\rm H}_{2}^{+}$ excited by one-cycle laser pulses 
 at $\lambda_{l}=150$ and 300~nm is presented in Figure~3 on the timescale of 3~fs 
 (left and right panel, respectively). 
 It is seen from Figures~3(a), 3(b), 3(d) and 3(e) that electron approximately follows 
 the laser field out-of-phase only during the first half-cycle, while during the second 
 half-cycle of the laser pulse, the expectation value $\langle z \rangle$ may change  
 even in-phase with the laser field. 
 Nevertheless, after the ends of one-cycle laser pulses, expectation values $\langle z \rangle$ 
 demonstrate rather regular oscillations with the period of $\tau_{\rm osc} \simeq 0.667$~fs corresponding 
 to the wavelength $\lambda_{\rm osc} \simeq 200$~nm and frequency $\omega_{\rm osc} \simeq 0.2278$~au. 
 
\begin{figure*}[t] 
\begin{center} 
\includegraphics*[width=0.90\textwidth]{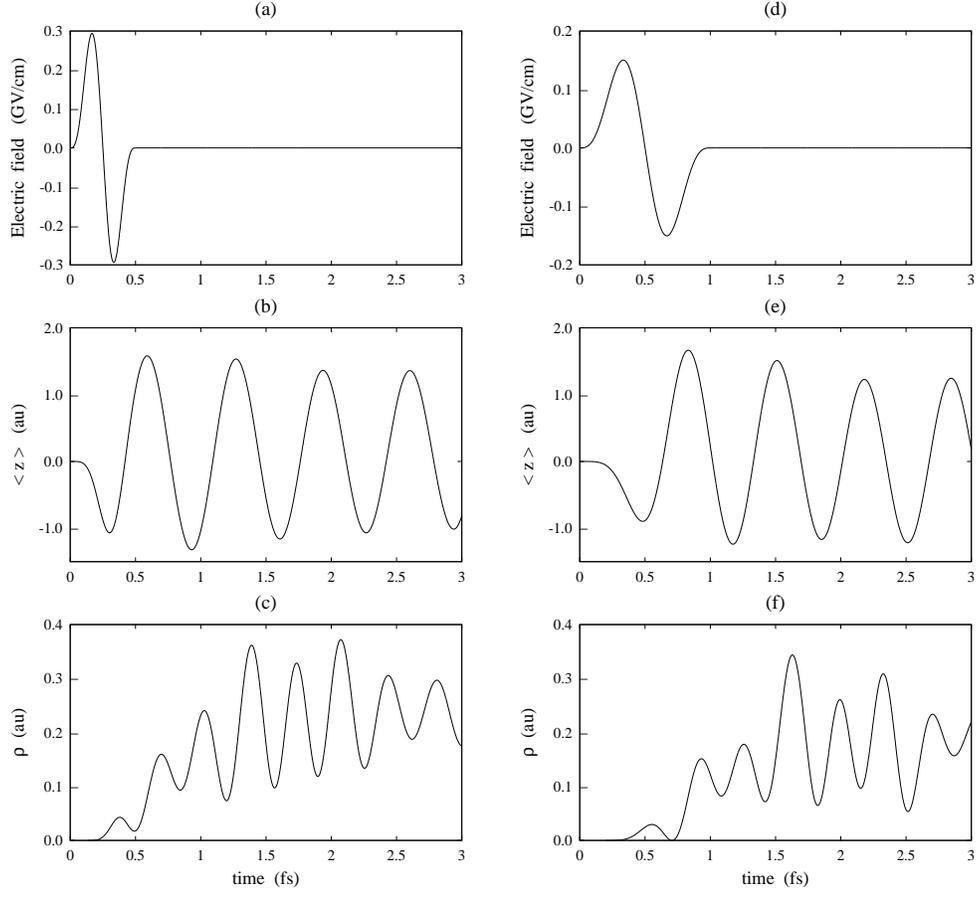} 
\end{center} 
 \caption{Quantum dynamics of ${\rm H}_{2}^{+}$ excited by the one-cycle laser pulses  
 at $\lambda_{l}=150$~nm (a-c) and $\lambda_{l}=300$~nm (d-f). 
 Parameters of the laser pulses: (a) ${\cal E}_{0}=0.088$~au, $\omega_{l}=0.30376$~au, 
 $t_{p}=0.5$~fs; (d) ${\cal E}_{0}=0.045$~au, $\omega_{l}=0.15188$~au, $t_{p}=1.0$~fs; 
 (b,e) time-dependent expectation values $\langle z \rangle$; 
 (c,f) time-dependent expectation values $\langle \rho \rangle$.} 
\end{figure*} 
 
 The transversal electron degree of freedom, $\rho$, is optically passive if the laser field  
 is aligned along the $z$ axis (Figure~1). 
 It can be excited therefore only due to the wave properties of electron. 
 The time-dependent expectation values $\langle \rho \rangle$ at laser wavelengths $\lambda_{l}=150$ and 300~nm 
 are presented in Figures~3(c) and (f), respectively. 
 It is seen that expectation values $\langle \rho \rangle$  
 start to increase at the end of the first optical half-cycle and demonstrates quite regular  
 (yet not harmonic) oscillations during the second optical half-cycle and after the ends of the one-cycle laser pulses. 
 Two major frequencies of $\rho$-oscillations can be distinguished at a close look.  
 Indeed, time intervals between two highest maxima of $\rho$-oscillations in Figures~3(c) and (f)  
 are $\tau^{\rm osc}_{1} \approx 0.667$~fs, corresponding to the frequency of $\omega^{\rho}_{1} \approx \omega_{\rm osc}$, 
 or the wavelength of $\lambda^{\rho}_{1} \approx 200$~nm.  
 On the other hand, time intervals between the neighboring maxima of $\rho$-oscillations in Figures~3(c) and (f), 
 are $\tau^{\rm osc}_{2} \approx 0.334$~fs, corresponding to the frequency of $\omega^{\rho}_{2} \approx 2\omega_{\rm osc}$, 
 or the wavelength of $\lambda^{\rho}_{2} \approx 100$~nm.  
 
 While the appearance of electronic $\rho$-oscillations with the frequency $\omega^{\rho}_{1} \approx \omega_{\rm osc}$  
 being very close to the frequency of electronic $z$-oscillations could be expected,  
 the ``frequency-doubling'' of electronic $\rho$-oscillations occurring at $\omega^{\rho}_{2} \approx 2\omega_{\rm osc}$ 
 needs more explanations. 
 Such a frequency-doubling of electronic $z$-oscillations by its $\rho$-oscillations occurring at  
 $\omega^{\rho}_{2} \approx 2\omega_{\rm osc}$ was explained in our previous works 
 \cite{Paramon:05cpl.HH-HD,%
 Paramon:07cp.HH-HD-Muon} 
 as follows.  
 During electronic oscillations along the $z$ axis, the electronic density is substantially delocalized  
 also in the transversal, $\rho$ direction, due to the wave properties of the electron. 
 This takes place at every turning point of electronic $z$-oscillations, i.e., twice per every cycle  
 of electronic $z$-oscillations, giving rise to excitation of ${\rm H}_{2}^{+}$ along the $\rho$ axis 
 in a stepwise manner. Similar frequency-doubling of muonic oscillation also takes place in muonic 
 $dd\mu$ and $dt\mu$ molecules 
 excited by super-intense soft X-ray laser pulses at the wavelength of $\lambda_{l}=5$~nm 
 on the attosecond timescale 
 \cite{BandrParamon:ijmpE.2014.muon}. 
 
 The characteristic feature of the post-laser-field electronic $z$-oscillations in ${\rm H}_{2}^{+}$ excited  
 by one-cycle laser pulses is their asymmetry with respect to $\langle z \rangle =0$ [Figures~3(b) and (e)]. 
 This is especially evident on a longer timescale in comparison with smoothly shaped oscillations of  
 expectation values $\langle - \partial V/\partial z \rangle$, as shown in Figure~4 on the timescale of 12~fs. 
 The left panel presents expectation values $\langle z \rangle$ at the laser wavelength $\lambda_{l}=150$~nm (a)  
 and $\lambda_{l}=300$~nm (b). 
 The right panel presents expectation values $\langle - \partial V/\partial z \rangle$  
 at $\lambda_{l}=150$~nm (c) and $\lambda_{l}=300$~nm (d). 
 Post-field oscillations of $\langle - \partial V/\partial z \rangle$ occur with the same frequency,  
 $\omega_{\rm osc} \simeq 0.2278$~au (corresponding to the wavelength $\lambda_{\rm osc} \simeq 200$~nm)   as those of $\langle z \rangle$.  
 The existence of coherent post-laser-field oscillations of $\langle z \rangle$ and $\langle - \partial V/\partial z \rangle$ 
 in ${\rm H}_{2}^{+}$ excited to the energy of $\langle E \rangle \simeq -0.515$~au by two-cycle laser pulses 
 was rationalized in detail in our recent work 
 \cite{ParKuehnBandr:jpcA.2016.H2p.post-field} 
 where approximately the same characteristic oscillation frequency  was calculated. 
Thus  in the present case of one-cycle laser pulses similar arguments hold true 
 to explain the appearance of post-laser-field electronic oscillations. 
 Note that the characteristic oscillation frequency corresponding to $\lambda_{\rm osc} \approx 200$~nm 
 is in line 
 with the results presented in Figure~2 evidencing the most efficient excitation of ${\rm H}_{2}^{+}$ to  
 $\langle E \rangle \simeq -0.515$~au at the laser wavelength of $\lambda_{l} \approx 200$~nm. 
 Also note that asymmetry of time-dependent expectation values $\langle z \rangle$ with respect to $\langle z \rangle =0$ 
 implies the time-dependent polarization of ${\rm H}_{2}^{+}$ after its excitation by one-cycle laser pulses. 
\begin{figure*}[t] 
\begin{center} 
\includegraphics*[width=0.90\textwidth]{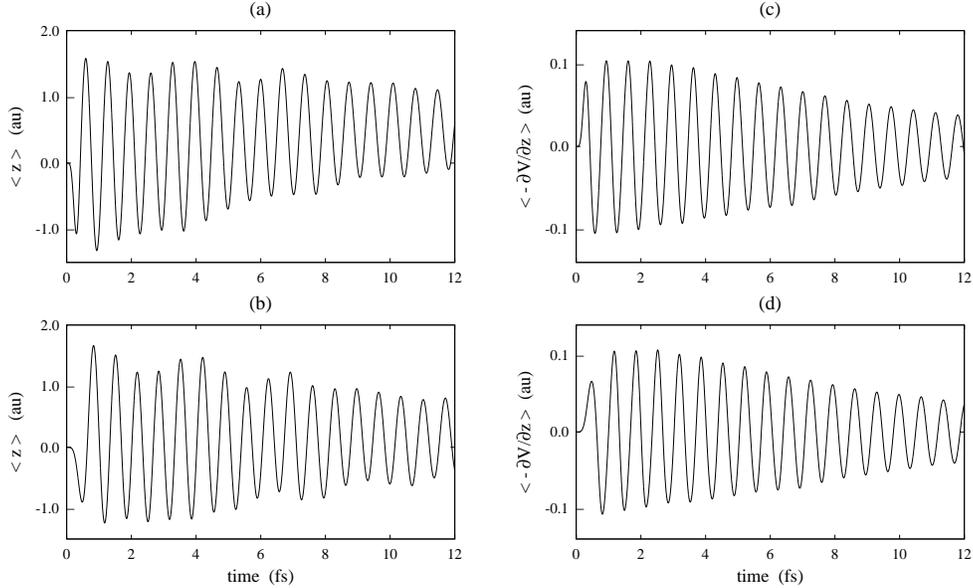} 
\end{center} 
 \caption{Quantum dynamics of ${\rm H}_{2}^{+}$ excited by the one-cycle laser pulses  
 at $\lambda_{l}=150$~nm (a,c) and $\lambda_{l}=300$~nm (b,d) followed by coherent post-laser-field electronic oscillations: 
 (a,b) time-dependent expectation values $\langle z \rangle$; 
 (c,d) time-dependent expectation values $\langle - \partial V/\partial z \rangle$. 
 Parameters of the laser pulses are as in Figure~3.}  
\end{figure*}

\subsection{Ionization, dissociation and post-laser-field electronic oscillations of  ${\rm H}_{2}^{+}$  on a long timescale} 
 
 Numerical simulations of the laser-driven quantum dynamics and subsequent free evolution of ${\rm H}_{2}^{+}$ 
 were performed as long as ionization and dissociation probabilities were small enough, not more than about 2\%. 
 
\begin{figure}[h] 
\begin{center} 
\includegraphics*[width=0.53\textwidth]{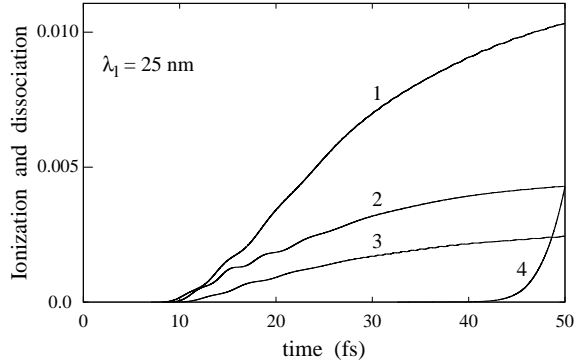} 
\end{center} 
\caption{Ionization and dissociation probabilities of ${\rm H}_{2}^{+}$ excited by one-cycle laser pulse 
 at the wavelength $\lambda_{l}=25$~nm.  
 Curve~1 --- ionization probability for the outer end of the $\rho $ axis;  
 curves~2 and 3 --- ionization probabilities for the negative and the positive direction of the $z$ axis, respectively;  
 curve~4 --- dissociation probability.  
 Parameters of the laser pulse: ${\cal E}_{0}=1.18$~au, $\omega_{l}=1.8225$~au, $t_{p} \simeq 0.0833$~fs.} 
\end{figure} 
 
 The time-dependent ionization and dissociation probabilities are presented in Figure~5 on the timescale of 50~fs 
 for the case when ${\rm H}_{2}^{+}$ is excited by one-cycle laser pulse at the laser wavelength $\lambda_{l}=25$~nm. 
 Note that ionization probabilities have been calculated with the respective time- and space-integrated outgoing fluxes 
 separately for the negative and the positive direction of the $z$ axis (curves~2 and 3, respectively) 
 as well as for the outer end of the $\rho $ axis (curve~1).  
 The dissociation probability (curve~4) has been similarly calculated for the nuclear coordinate $R$. 
 
 Since the optimal laser-pulse amplitude required to prepare ${\rm H}_{2}^{+}$ at the energy of  
 $\langle E \rangle \approx -0.515$~au 
 at the laser wavelength $\lambda_{l}=25$~nm is substantially larger than at larger wavelengths, 
 both dissociation and ionization probabilities presented in Figure~5 are the largest obtained in this work. 
 It is seen from Figure~5 that the ionization probability for the outer end of the $\rho $ axis is larger 
 than those for both negative and positive direction of the $z$ axis.  
 This can be explained by the aforementioned fact that the electronic density is delocalized in the $\rho$ direction 
 twice per every cycle of electronic $z$-oscillations, giving rise to the frequency-doubling of electronic 
 $\rho$-oscillations as compared to electronic $z$-oscillations. 
 It is also seen from Figure~5 that ionization starts much prior to dissociation, therefore `ionizative dissociation' occurs. 
 Indeed, the decrease of the electron density between the two protons due to ionization disturbs the initial equilibrium 
 configuration of ${\rm H}_{2}^{+}$ and thus allows the Coulombic repulsion of the protons to act more efficiently 
 resulting in the elongation of the internuclear distance in ${\rm H}_{2}^{+}$ and its subsequent dissociation. 
 
\begin{figure}[h] 
\begin{center} 
\includegraphics*[width=0.53\textwidth]{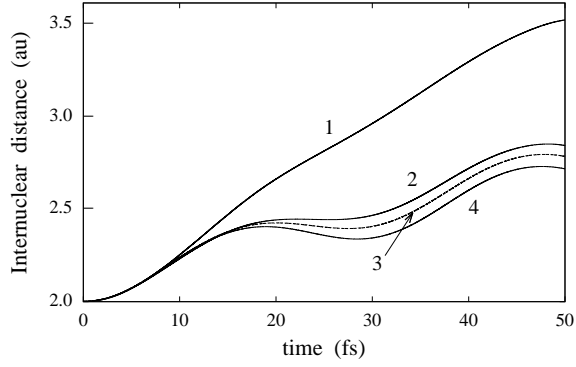} 
\end{center} 
 \caption{The time-dependent expectation values $\langle R \rangle$ 
 for ${\rm H}_{2}^{+}$ excited by the one-cycle laser pulses 
 at $\lambda_{l}=25$~nm (curve~1, ${\cal E}_{0}=1.18$~au, $\omega_{l}=1.8225$~au, $t_{p} \simeq 0.0833$~fs),  
 $\lambda_{l}=50$~nm (curve~2, ${\cal E}_{0}=0.625$~au, $\omega_{l}=0.91127$~au, $t_{p} \simeq 0.167$~fs),  
 $\lambda_{l}=100$~nm (curve~3, ${\cal E}_{0}=0.174$~au, $\omega_{l}=0.45563$~au, $t_{p} \simeq 0.333$~fs) 
 and $\lambda_{l}=200$~nm (curve~4, ${\cal E}_{0}=0.06$~au, $\omega_{l}=0.22782$~au, $t_{p} \simeq 0.667$~fs).} 
\end{figure} 
 
 The time-dependent expectation values $\langle R \rangle$ of internuclear distances in ${\rm H}_{2}^{+}$  
 are presented in Figure~6 on the timescale of 50~fs for the laser wavelengths of  
 $\lambda_{l}=25$~nm (curve~1), $\lambda_{l}=50$~nm (curve~2), $\lambda_{l}=100$~nm (curve~3) and 
 $\lambda_{l}=200$~nm (curve~4). 
 It is seen from Figure~6 that at $\lambda_{l}=50$, 100 and 200~nm (curves~2, 3 and 4, respectively),  
 the time-dependent internuclear distances $R$ 
 behave very similar to each other, all demonstrating local maxima at $t \approx 17$~fs and at $t \approx 47$~fs  
 (elongation of the bond) as well as local minima at $t \approx 32$~fs (contraction of the bond length). 
 Curve~1, corresponding to the case of $\lambda_{l}=25$~nm, looks at a first glance as an exception due to much more 
 substantial increase of the bond length caused by a comparatively strong laser field applied to prepare ${\rm H}_{2}^{+}$  
 close to its dissociation threshold and, therefore, more efficient ionizative dissociation of ${\rm H}_{2}^{+}$  
 resulting in a more substantial overall bond elongation.  
 Nevertheless, both elongation and contraction of the internuclear distance 
 can be seen in curve~1 as well.  
 Since the nuclear motion in the symmetric ${\rm H}_{2}^{+}$ molecule is activated only by the electronic motion induced 
 by the laser field along the $z$ axis, it is interesting to find a correlation between the nuclear motion and post-laser-field 
 electronic $z$-oscillations. 
 
\begin{figure*}[h] 
\begin{center} 
\includegraphics*[width=0.90\textwidth]{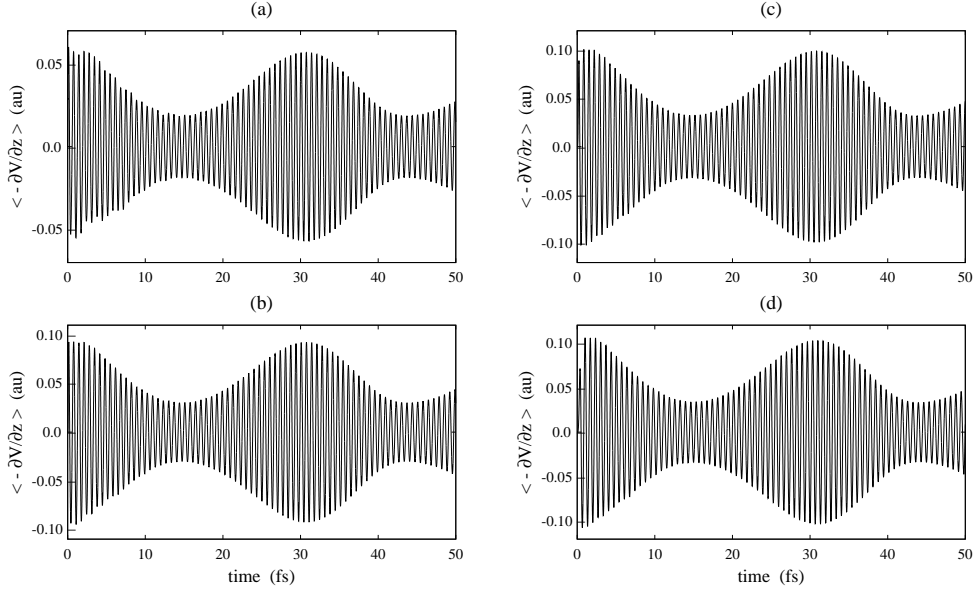} 
\end{center} 
 \caption{The time-dependent expectation values $\langle - \partial V/\partial z \rangle$ for ${\rm H}_{2}^{+}$  
 excited by the one-cycle laser pulses at $\lambda_{l}=25$~nm (a), $\lambda_{l}=50$~nm (b), 
 $\lambda_{l}=100$~nm (c) and $\lambda_{l}=200$~nm (d). 
 Other parameters of the laser pulses are as in Figure~6.} 
\end{figure*} 
 
 In Figure~7 post-laser-field electronic oscillations are presented by the time-dependent expectation values  
 $\langle - \partial V/\partial z \rangle$ for the same laser wavelengths $\lambda_{l}$ as in Figure~6. 
 It is clearly seen from Figure~7 that local minima of the Coulomb force $\langle - \partial V/\partial z \rangle$ 
 correspond to local maxima of the bond length $\langle R \rangle$ of ${\rm H}_{2}^{+}$ (elongation of the bond),  
 while the local maximum of the Coulomb force $\langle - \partial V/\partial z \rangle$ at $t \approx 32$~fs corresponds 
 to the local minimum of the bond length (contraction of the bond length). 
 We can conclude therefore, that periodic elongation-contraction of the bond of ${\rm H}_{2}^{+}$ (Figure~6) 
 is controlled by compressing-expanding electron acceleration along the $z$ axis (Figure~7) which takes place with 
 the period of $\tau_{\rm shp} \approx 30$~fs corresponding to the frequency  
 $\omega_{\rm shp}=2\pi/\tau_{\rm shp}$ of shaped post-laser-field oscillations occurring with the carrier oscillation 
 frequency $\omega_{\rm osc} \simeq 0.2278$~au (corresponding to the wavelength of $\lambda_{\rm osc} \simeq 200$~nm). 
 Similar electron-nuclei correlations were found in our recent work  
 \cite{ParKuehnBandr:jpcA.2016.H2p.post-field} 
 where two-cycle laser pulses were used to excite ${\rm H}_{2}^{+}$ and a more detailed explanation is given for 
 post-laser-field electron-nuclei correlations in terms of below-resonance vibrational frequency 
 \cite{Paramon:07cp.HH-HD-Muon,%
 ParamonKuhn:2012jpca.H2p.below-res} 
 and for the existence of characteristic oscillation frequency $\omega_{\rm osc}$  
 (corresponding to $\lambda_{\rm osc} \simeq 200$~nm)  
 in terms of a continuum state $\Psi_{C}$ prepared by the laser pulses. 
 
\begin{figure}[h] 
\begin{center} 
\includegraphics*[width=0.53\textwidth]{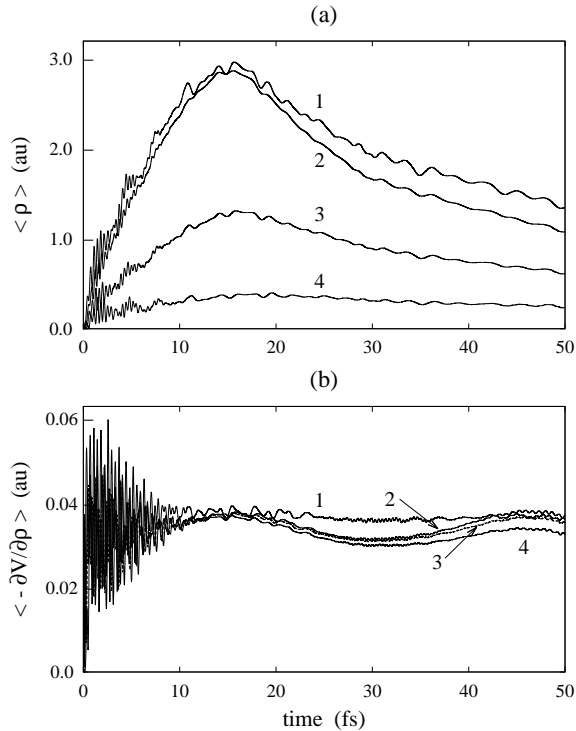} 
\end{center} 
 \caption{The time-dependent expectation values $\langle \rho \rangle$ (a)  
 and $\langle - \partial V/\partial \rho \rangle$ (b) for ${\rm H}_{2}^{+}$  
 excited by the one-cycle laser pulses at $\lambda_{l}=25$~nm (curves~1), $\lambda_{l}=50$~nm (curves~2), 
 $\lambda_{l}=100$~nm (curves~3) and $\lambda_{l}=200$~nm (curves~4). 
 Other parameters of the laser pulses are as in Figure~6.} 
\end{figure} 
 
 Finally, to complete this section, we present in Figure~8 the time-dependent expectation values $\langle \rho \rangle$ 
 [Figure~8(a)] and $\langle - \partial V/\partial \rho \rangle$ [Figure~8(b)] calculated on the long timescale of 50~fs 
 at the laser wavelengths of $\lambda_{l}=25$, 50, 100 and 200~nm. 
  
 It is seen from Figures~8(a) and 8(b) that expectation values $\langle \rho \rangle$  
 and $\langle - \partial V/\partial \rho \rangle$ first demonstrate fast oscillations on the time interval of about 
 $t>7-10$~fs. Again, two major frequencies of electronic $\rho$-oscillations can be distinguished at a close look:  
 $\omega^{\rho}_{1} \approx \omega_{\rm osc}$ ($\lambda^{\rho}_{1} \approx 200$~nm) and  
 $\omega^{\rho}_{2} \approx 2\omega_{\rm osc}$ ($\lambda^{\rho}_{2} \approx 100$~nm). 
 The frequency-doubling of electronic $\rho$-oscillations has been already described in Section~3 and presented 
 in Figures~3(c) and 3(f) therein. 
 
 Afterwards, at $t>7-9$~fs, expectation values $\langle \rho \rangle$ demonstrate a rather smooth behavior, 
 with the electron excursion along the transversal $\rho$ coordinate being quite strongly dependent of the laser pulse  
 amplitude used to excite ${\rm H}_{2}^{+}$ initially along the $z$ axis. 
 It is also seen from Figure~8(b) that there exists a nice correlation between the time-dependent 
 $\langle - \partial V/\partial \rho \rangle$ values and the internuclear distances $\langle R \rangle$ of Figure~6. 
 Indeed, at $\lambda_{l}=50$, 100 and 200~nm [curves~2, 3 and 4, respectively, in both Figure~6 and Figure~8(b)], 
 local maxima of both $\langle - \partial V/\partial \rho \rangle$ and $\langle R \rangle$ occur at $t \approx 17$~fs  
 and at $t \approx 47$~fs, while their local minima occur at $t \approx 32$~fs.  
 Since no frequency-doubling of post-laser-field electronic $\rho$-oscillations (as compared to $z$-oscillations) takes place,  
 we conclude that the low-frequency oscillations of $\langle - \partial V/\partial \rho \rangle$ at $t>10$~fs  
 are induced by the periodic elongation-contraction of the bond length in ${\rm H}_{2}^{+}$ (Figure~6), rather 
 than by post-laser-field electronic $z$-oscillations (Figure~7).  
 Again, since the periodic elongation-contraction of the bond length in ${\rm H}_{2}^{+}$ is very small at $\lambda_{l}=25$~nm 
 (curve~1 in Figure~6), local maxima and minima of $\langle - \partial V/\partial \rho \rangle$ are not seen  
 at $\lambda_{l}=25$~nm as well [curve~1 in Figure~8(b)]. 
  
 \subsection{Power spectra generated by post-laser-field electronic oscillations in ${\rm H}_{2}^{+}$} 
  
 In this section we present the power spectra generated by post-laser-field electronic motion  
 calculated in the acceleration form, $A_{z}(\omega)$ and $A_{\rho}(\omega)$, 
 for the electron coordinates $z$ and $\rho$, respectively. 
 
 It is straightforward to show with the Ehrenfest's theorem, 
 that the acceleration of the expectation value $\langle z \rangle $ can be written in the following form: 
 \begin{equation} 
  \frac{d^{2}}{dt^{2}} \langle z \rangle =  
  - \frac{1}{m_{\rm e}}\bigl[\langle \partial V/\partial z \rangle + {\cal E}(t)\bigr]\, . 
 \label{z-acceleration-long} 
 \end{equation} 
 Since the applied laser field aligned along the $z$ axis does not excite the transversal $\rho $ degree of freedom directly,   
 its excitation can occur only due to the wave properties of electron. 
 Therefore, the electric-field term ${\cal E}(t)$ does not appear in 
 the equation for the acceleration of the expectation value $\langle \rho \rangle$ at all. 
 It can also be shown, by making use of Ehrenfest's theorem, that the acceleration of  
 the expectation value $\langle \rho \rangle $ reads 
 \begin{equation} 
  \frac{d^{2}}{dt^{2}} \langle \rho \rangle =  
   - \frac{1}{m_{\rm e}}\langle \partial V/\partial \rho \rangle . 
 \label{Rho-acceleration}  
  \end{equation} 
 
The power spectrum $S(\omega)$ of any time-dependent expectation value $\langle S(t) \rangle $ is defined  
 by the squared modulus of the Fourier transform: 
 \begin{equation} 
  S(\omega)=\bigg|\int_{0}^{t_{f}} \langle S(t) \rangle \exp(-i\omega t) dt \bigg|^{2}, 
  \label{S-FT} 
  \end{equation} 
 where 
 \begin{equation} 
 \langle S(t) \rangle = \langle \Psi(t)|S|\Psi(t) \rangle. 
 \label{S-definition} 
 \end{equation} 
 
 In the case under consideration, the time-dependent expectation value $\langle S(t) \rangle $ in Equation~(\ref{S-FT})  
 will stand accordingly for $d^{2}\langle z \rangle /dt^{2}$ defined by Equation~(\ref{z-acceleration-long}),  
 or for $d^{2}\langle \rho \rangle /dt^{2}$ defined by Equation~(\ref{Rho-acceleration}). 
 In the given above definitions we took into account  
 that the power spectra defined by Equations~(\ref{S-FT}) and (\ref{S-definition}) 
 do not depend on the sign of $S$. 
 
 \subsubsection{Generation of lower-order harmonics at $\lambda_{l}<200$~nm} 
 
 As it was discussed earlier (Section~3.1), power spectra resulting from excitation of ${\rm H}_{2}^{+}$ 
 close to its dissociation threshold at various laser wavelengths $\lambda_{l} \neq 200$~nm  
 might always generate harmonics corresponding to $\lambda \simeq 200$~nm due to the most efficient excitation 
 of ${\rm H}_{2}^{+}$ at this wavelength (Figure~2). 
 Therefore, at $\lambda_{l} < 200$~nm, generation of lower-order harmonics with respect to electronic $z$-motion 
 can be expected. 
 
 In Figure~9, power spectra in the acceleration form, $A_{z}(\omega)$ and $A_{\rho}(\omega)$,  
 generated due to the laser-initiated electron 
 motion along the $z$ coordinate, are presented for the case when ${\rm H}_{2}^{+}$ is excited by one-cycle 
 laser pulse at the laser wavelength of $\lambda_{l}=25$~nm. 
 
\begin{figure}[h] 
\begin{center} 
\includegraphics*[width=0.53\textwidth]{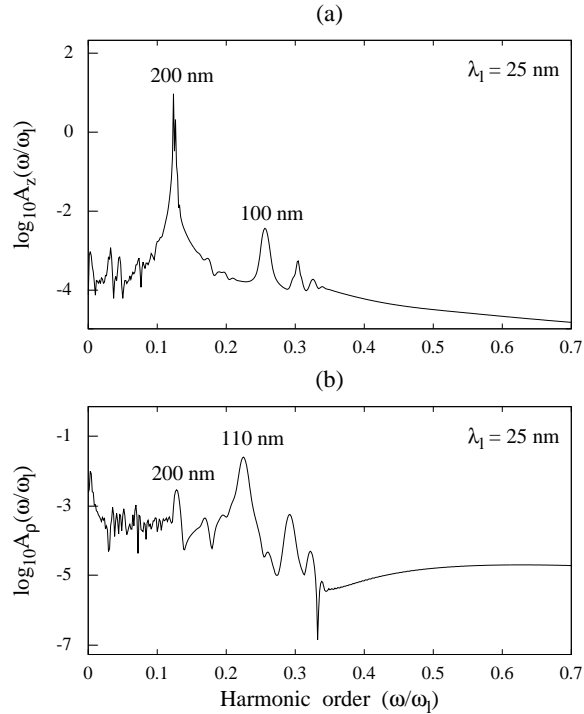} 
\end{center} 
\caption{Generation of lower-order harmonics in ${\rm H}_{2}^{+}$ excited by the one-cycle laser pulse 
 at $\lambda_{l}=25$~nm (${\cal E}_{0}=1.18$~au, $\omega_{l}=1.8225$~au, $t_{p} \simeq 0.0833$~fs): 
 (a) power spectrum $A_{z}(\omega)$ generated by the optically active $z$ degree of freedom; 
 (b) power spectrum $A_{\rho}(\omega)$ generated by the optically passive, transversal $\rho$ degree of freedom.} 
\end{figure} 
 
 It is seen from Figure~9(a) that the strongest lower-order harmonic of the $A_{z}$ spectrum corresponds, as it was expected, 
 to the wavelength of $\lambda \simeq 200$~nm, while a weaker lower-order harmonic corresponds  
 to $\lambda \simeq 100$~nm. 
 In the power spectrum $A_{\rho}$ generated by the transversal $\rho$ degree of freedom, Figure~9(b), 
 the strongest lower-order harmonic at $\lambda \simeq 110$~nm corresponds 
 to the doubled frequency of $\rho$-oscillations $\omega^{\rho}_{2} \approx 2\omega_{\rm osc}$  
 (where $\omega_{\rm osc}$ corresponds to the wavelength of $\lambda \simeq 200$~nm). 
 It is also seen from Figure~9(b) that the lower order harmonic $\omega^{\rho}_{1} \approx \omega_{\rm osc}$ 
 corresponding to $\lambda \simeq 200$~nm occurs in the power spectrum $A_{\rho}(\omega)$ as well. 

\begin{figure}[t] 
\begin{center} 
\includegraphics*[width=0.53\textwidth]{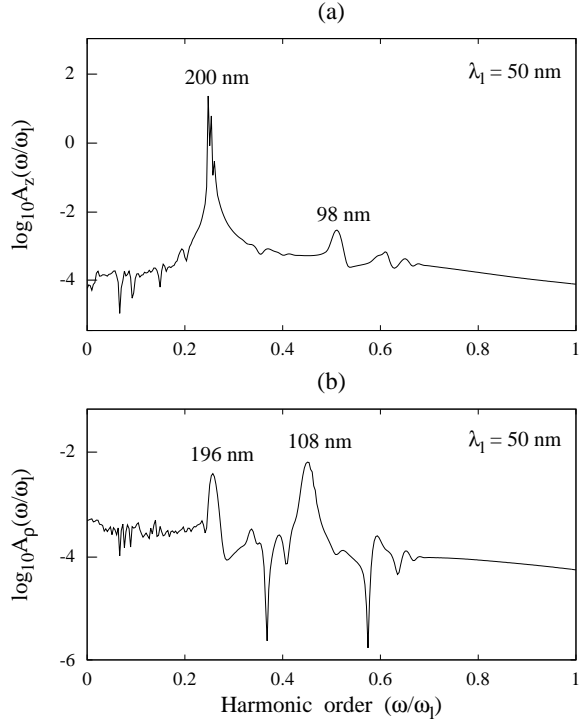} 
\end{center} 
\caption{Generation of lower-order harmonics in ${\rm H}_{2}^{+}$ excited by the one-cycle laser pulse 
 at $\lambda_{l}=50$~nm (${\cal E}_{0}=0.625$~au, $\omega_{l}=0.91127$~au, $t_{p} \simeq 0.167$~fs): 
 (a) power spectrum $A_{z}(\omega)$ generated by the optically active $z$ degree of freedom; 
 (b) power spectrum $A_{\rho}(\omega)$ generated by the optically passive, transversal $\rho$ degree of freedom.} 
\end{figure} 
 
 In Figure~10, power spectra $A_{z}(\omega)$ and $A_{\rho}(\omega)$,  
 are presented for the case when ${\rm H}_{2}^{+}$ is excited by one-cycle 
 laser pulse at the laser wavelength of $\lambda_{l}=50$~nm. 
 Again, as it is seen from Figure~10(a), the strongest lower-order harmonic of the $A_{z}$ spectrum corresponds 
 to the wavelength of $\lambda \simeq 200$~nm,  
 while the second, much weaker, lower-order harmonic corresponds to $\lambda \simeq 98$~nm 
 (i.e., it is very close to $\lambda = 100$~nm). 
 In the power spectrum $A_{\rho}$ of the transversal $\rho$ degree of freedom, Figure~10(b), 
 the strongest lower-order harmonic at $\lambda \simeq 108$~nm corresponds   
 to the doubled frequency of $\rho$-oscillations $\omega^{\rho}_{2} \approx 2\omega_{\rm osc}$, 
 while a weaker lower order harmonic at $\lambda \simeq 196$~nm corresponds to $\omega^{\rho}_{1} \approx \omega_{\rm osc}$. 
 
 Power spectra $A_{z}(\omega)$ and $A_{\rho}(\omega)$ for the case of the laser wavelength $\lambda_{l}=100$~nm 
 are shown in Figures~11(a) and (b), respectively.  
 The strongest lower-order harmonic of the $A_{z}$ spectrum generated by the optically active $z$ degree of freedom 
 corresponds to the wavelength of $\lambda \simeq 200$~nm, 
 while a comparatively very weak harmonic corresponds to $\lambda \simeq 100$~nm.  
 We can conclude therefore that at the laser wavelength $\lambda_{l}=100$~nm one lower-order and one ``identical'' 
 harmonic are generated by the optically active $z$ degree of freedom of ${\rm H}_{2}^{+}$. 
 
 In the power spectrum $A_{\rho}$ generated by the optically passive $\rho$ degree of freedom, Figure~11(b), 
 a new feature can be observed as well. 
 Indeed, while the strongest lower-order harmonic at $\lambda \simeq 110$~nm corresponds  
 to $\omega^{\rho}_{2} \approx 2\omega_{\rm osc}$, 
 and a weaker lower order harmonic at $\lambda \simeq 200$~nm corresponds to $\omega^{\rho}_{1} \approx \omega_{\rm osc}$, 
 the other harmonic corresponding to $\omega^{\rho}_{2} \approx 2\omega_{\rm osc}$,  
at $\lambda \simeq 85$~nm, is the higher-order harmonic with respect to the laser wavelength  
 $\lambda_{l}=100$~nm used to excite ${\rm H}_{2}^{+}$. 
 We can conclude therefore, that the laser wavelength of $\lambda_{l}=100$~nm manifests itself as a beginning 
 of the appearance of the higher-order harmonics, at least in the power spectra $A_{\rho}(\omega)$  
 generated by the optically passive, transversal $\rho$ degree of freedom. 
 The physical reason behind this feature is the above described frequency-doubling of the electronic $\rho$-oscillations, 
 with respect to the laser-initiated electronic $z$-oscillations, caused by the wave-properties of an electron. 
 
\begin{figure}[t] 
\begin{center} 
\includegraphics*[width=0.53\textwidth]{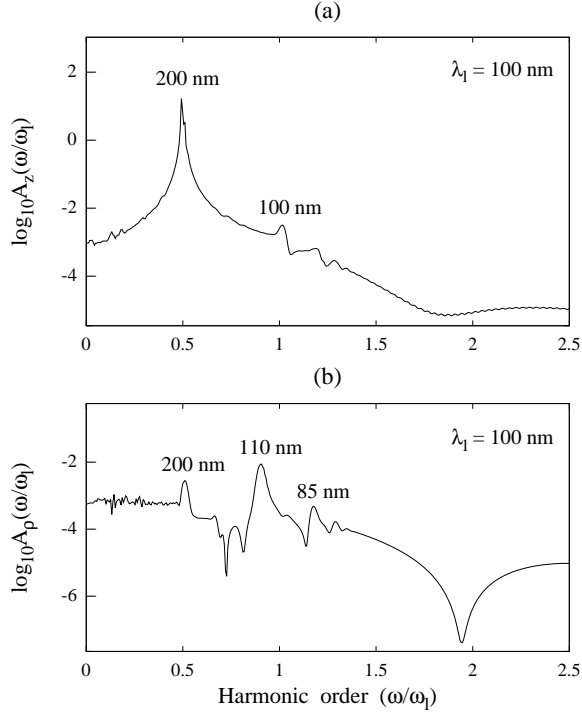} 
\end{center} 
\caption{Generation of lower- and higher-order harmonics in ${\rm H}_{2}^{+}$ excited by the one-cycle laser pulse 
 at $\lambda_{l}=100$~nm (${\cal E}_{0}=0.174$~au, $\omega_{l}=0.45563$~au, $t_{p} \simeq 0.333$~fs): 
 (a) power spectrum $A_{z}(\omega)$ generated by the optically active $z$ degree of freedom; 
 (b) power spectrum $A_{\rho}(\omega)$ generated by the optically passive, transversal $\rho$ degree of freedom.} 
\end{figure} 
 
\begin{figure}[t] 
\begin{center} 
\includegraphics*[width=0.53\textwidth]{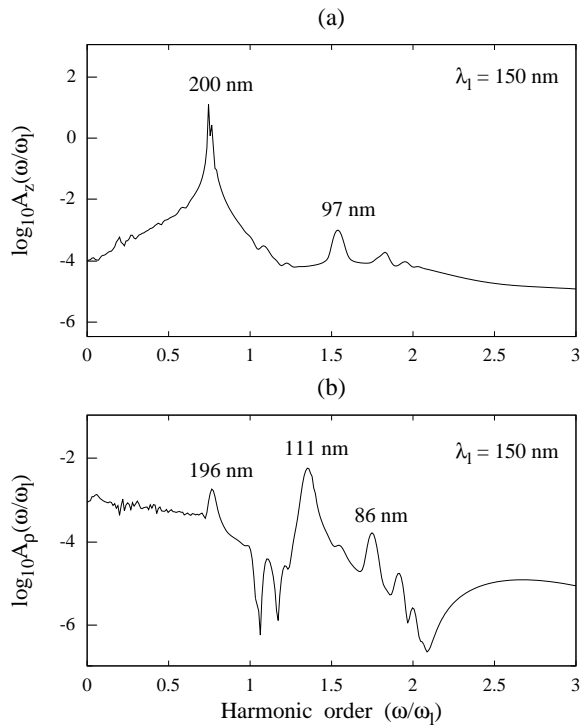} 
\end{center} 
\caption{Generation of lower- and higher-order harmonics in ${\rm H}_{2}^{+}$ excited by the one-cycle laser pulse 
 at $\lambda_{l}=150$~nm (${\cal E}_{0}=0.088$~au, $\omega_{l}=0.30376$~au, $t_{p} \simeq 0.5$~fs): 
 (a) power spectrum $A_{z}(\omega)$ generated by the optically active $z$ degree of freedom; 
 (b) power spectrum $A_{\rho}(\omega)$ generated by the optically passive, transversal $\rho$ degree of freedom.} 
\end{figure} 
 
 The situation is developing further at a larger wavelength, $\lambda_{l}=150$~nm, approaching the most efficient 
 for the excitation of ${\rm H}_{2}^{+}$ close to its dissociation threshold laser wavelength of $\lambda_{l}=200$~nm, 
 as depicted in Figure~2. 
 
 Power spectra $A_{z}(\omega)$ and $A_{\rho}(\omega)$ are presented in Figure~12 for the case when ${\rm H}_{2}^{+}$  
 is excited by one-cycle laser pulse at the laser wavelength of $\lambda_{l}=150$~nm. 
 It is clearly seen from Figure~12(a) for the $A_{z}(\omega)$ spectrum that the strongest harmonic at $\lambda \simeq 200$~nm 
 is still the lower-order harmonic with respect to the laser wavelength $\lambda_{l}=150$~nm used to excite ${\rm H}_{2}^{+}$, 
 while the other, a weak harmonic at $\lambda \simeq 97$~nm, is the higher-order harmonic. 
 
 In the power spectrum $A_{\rho}(\omega)$ of Figure~12(b) for the optically passive $\rho$ degree of freedom, 
 the lower-order harmonic at $\lambda \simeq 196$~nm corresponding to $\omega^{\rho}_{1} \approx \omega_{\rm osc}$  
 is not the strongest one anymore. The strongest harmonic at $\lambda \simeq 111$~nm is a higher-order harmonic,  
 as well as a weaker one at $\lambda \simeq 86$~nm, both corresponding to $\omega^{\rho}_{2} \approx 2\omega_{\rm osc}$ 
the accordance with the frequency-doubling of the electronic $\rho$-oscillations. 
 
 We can conclude therefore from the results presented in this section (Figures~9 through 12)  
 that when the laser wavelength $\lambda_{l}$ increases from small values 
 of about 25~nm and approaches 200~nm, generation of higher-order harmonics starts,  
 due to the frequency-doubling of the electronic $\rho$-oscillations, from the optically passive $\rho$ degree of freedom 
 at $\lambda_{l}=100$~nm [Figure~11(b)] and develops such that both $A_{\rho}(\omega)$ and $A_{z}(\omega)$ power spectra 
 have higher-order harmonics at $\lambda_{l}=150$~nm (Figure~12). 
 
 \subsubsection{Generation of higher-order harmonics at $\lambda_{l} \geq 200$~nm} 
 
 As it was already discussed in Section~3.1 and depicted in Figure~2 therein, there are two different domains 
 of the laser wavelength $\lambda_{l}$ characterizing the efficiency of excitation of ${\rm H}_{2}^{+}$  
 to the energy close to its dissociation threshold by one-cycle and two-cycle laser pulses. 
 The domain of $\lambda_{l} \geq 200$~nm to be considered in this section corresponds to a very small change 
 of the optimal laser pulse amplitude ${\cal E}_{0}$ required to excite ${\rm H}_{2}^{+}$ from its ground state 
 to the energy of $\langle E \rangle \approx -0.515$~au. 
 Nevertheless, since $\lambda_{l} \approx 200$~nm is still the most efficient laser wavelength (see Figure~2), 
 we can expect that power spectra resulting from excitation of ${\rm H}_{2}^{+}$ at various laser wavelengths  
 $\lambda_{l} \geq 200$~nm would contain harmonics corresponding to $\lambda \approx 200$~nm. 
 
\begin{figure}[t] 
\begin{center} 
\includegraphics*[width=0.53\textwidth]{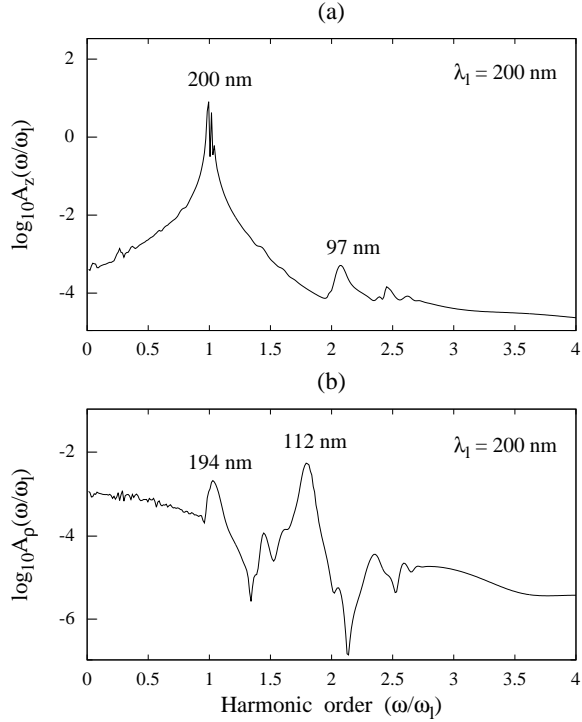} 
\end{center} 
\caption{Generation of higher-order harmonics in ${\rm H}_{2}^{+}$ excited by the one-cycle laser pulse 
 at $\lambda_{l}=200$~nm (${\cal E}_{0}=0.06$~au, $\omega_{l}=0.22782$~au, $t_{p} \simeq 0.667$~fs): 
 (a) power spectrum $A_{z}(\omega)$ generated by the optically active $z$ degree of freedom; 
 (b) power spectrum $A_{\rho}(\omega)$ generated by the optically passive, transversal $\rho$ degree of freedom.} 
\end{figure} 
 
 In Figure~13, power spectra in the acceleration form, $A_{z}(\omega)$ and $A_{\rho}(\omega)$,  
 generated due to the laser-initiated electron 
 motion along the $z$ coordinate, are presented for the case when ${\rm H}_{2}^{+}$ is excited by one-cycle 
 laser pulse at the laser wavelength of $\lambda_{l}=200$~nm. 
   
 It is seen from Figure~13(a) that the strongest harmonic in the $A_{z}$ spectrum generated by the optically active 
 $z$ degree of freedom is the first, or identical harmonic corresponding to the wavelength of $\lambda \simeq 200$~nm, 
 while a much weaker, the second-order harmonic corresponds to $\lambda \simeq 97$~nm. 
 The appearance of the second-order harmonic in the power spectrum $A_{z}$ of the symmetric ${\rm H}_{2}^{+}$ molecule 
 is the specific feature of one-cycle and two-cycle 
 \cite{ParKuehnBandr:jpcA.2016.H2p.post-field} 
 laser pulses, because if laser pulses with many optical cycles are used to excite a symmetric molecule, 
 even harmonics should not appear in the power spectra generated by optically active degrees of freedom at all, 
 as suggested by the concept of inversion symmetry 
 \cite{Gross:prl.2001.HHG}. 
 
 In the power spectrum $A_{\rho}$ generated by the optically passive $\rho$ degree of freedom, Figure~13(b), 
 the strongest second-order harmonic at $\lambda \simeq 112$~nm corresponds 
 to the doubled frequency of $\rho$-oscillations $\omega^{\rho}_{2} \approx 2\omega_{\rm osc}$, 
 while a weaker, the first-order harmonic at $\lambda \simeq 194$~nm, corresponds to $\omega^{\rho}_{1} \approx \omega_{\rm osc}$. 
 
\begin{figure}[t] 
\begin{center} 
\includegraphics*[width=0.53\textwidth]{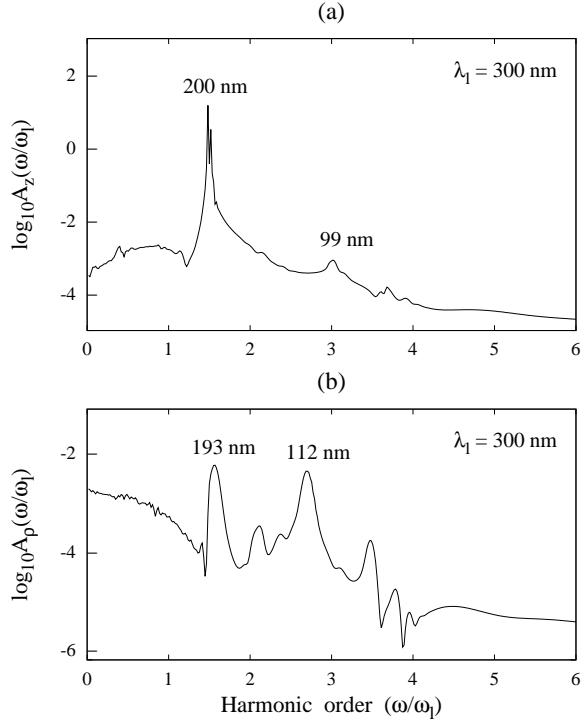} 
\end{center} 
\caption{Generation of higher-order harmonics in ${\rm H}_{2}^{+}$ excited by the one-cycle laser pulse 
 at $\lambda_{l}=300$~nm (${\cal E}_{0}=0.045$~au, $\omega_{l}=0.15188$~au, $t_{p} \simeq 1.0$~fs): 
 (a) power spectrum $A_{z}(\omega)$ generated by the optically active $z$ degree of freedom; 
 (b) power spectrum $A_{\rho}(\omega)$ generated by the optically passive, transversal $\rho$ degree of freedom.} 
\end{figure} 
 
 In Figure~14, power spectra $A_{z}(\omega)$ and $A_{\rho}(\omega)$ are presented for the case when ${\rm H}_{2}^{+}$  at the laser wavelength of $\lambda_{l}=300$~nm. 
 Again, as it is seen from Figure~14(a), the strongest harmonic in the $A_{z}$ spectrum generated by the optically active 
 $z$ degree of freedom corresponds to the wavelength of $\lambda \simeq 200$~nm.  
 The other higher-order harmonic in the power spectrum $A_{z}(\omega)$, that corresponding to $\lambda \simeq 99$~nm, 
 is (formally) the third-order harmonic with respect to the laser wavelength of $\lambda_{l}=300$~nm used to excite 
 ${\rm H}_{2}^{+}$. On the other hand, the higher-order harmonic at $\lambda \simeq 99$~nm is the second-order harmonic 
 with respect to the smallest-frequency harmonic at $\lambda \simeq 200$~nm appeared in the $A_{z}(\omega)$ power spectrum. 
 Such a new nomenclature of the harmonic order may often be very suitable, 
 because the smallest-frequency harmonics in all power spectra $A_{z}(\omega)$ generated by the optically-active $z$ degree  
 of freedom analyzed in this section above [see Figures~9(a) through 14(a)] are those corresponding to $\lambda \simeq 200$~nm. 
 The physical reason behind this feature is that $\lambda_{l} \simeq 200$~nm is the most efficient laser wavelength 
 with respect to excitation of ${\rm H}_{2}^{+}$ from the ground state to the energy close to its dissociation threshold 
 by one-cycle and two-cycle laser pulses, see Figure~2. 
 
 Since the smallest-frequency harmonics in all power spectra $A_{\rho}(\omega)$ generated by the optically-passive $\rho$ degree 
 of freedom [Figures~9(b) through 14(b)] also correspond to $\lambda_{l} \approx 200$~nm, a new nomenclature of the harmonic 
 order described above for $A_{z}(\omega)$ power spectra can be used to analyze power spectra $A_{\rho}(\omega)$ as well. 
 Indeed, it is seen from Figure~14(b) that the higher-order harmonic with the smallest-frequency is that  
 at $\lambda \simeq 193$~nm, while the higher-order harmonic at $\lambda \simeq 112$~nm is approximately the second-order 
 harmonic with respect to the smallest-frequency one (at $\lambda \simeq 193$~nm) and the third-order harmonic with respect 
 to the laser wavelength of $\lambda_{l}=300$~nm used to excite ${\rm H}_{2}^{+}$. 
 Needless to add that, as usual, harmonic at $\lambda \simeq 193$~nm corresponds to the frequency of  
 $\omega^{\rho}_{1} \approx \omega_{\rm osc}$, while harmonic at $\lambda \simeq 112$~nm corresponds to the doubled frequency  
 of $\rho$-oscillations $\omega^{\rho}_{2} \approx 2\omega_{\rm osc}$, where the oscillation frequency $\omega_{\rm osc}$ 
 corresponds to the wavelength $\lambda_{\rm osc} \approx 200$~nm. 
 
\begin{figure}[t] 
\begin{center} 
\includegraphics*[width=0.53\textwidth]{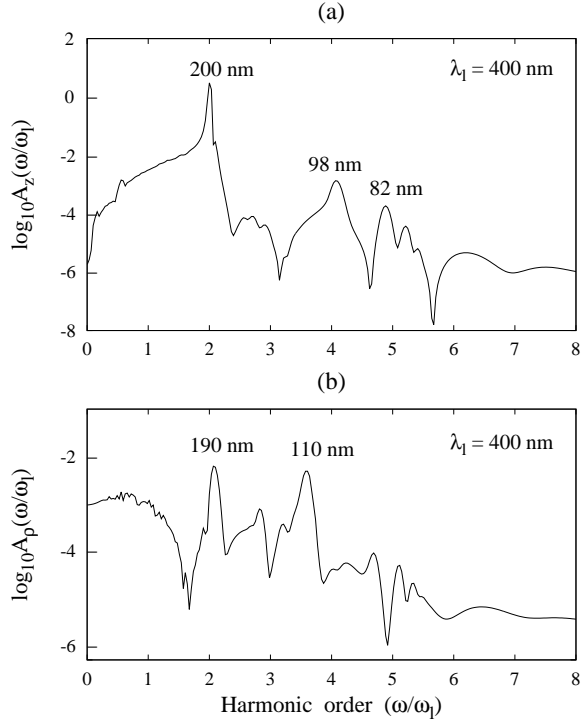} 
\end{center} 
\caption{Generation of higher-order harmonics in ${\rm H}_{2}^{+}$ excited by the one-cycle laser pulse 
 at $\lambda_{l}=400$~nm (${\cal E}_{0}=0.056$~au, $\omega_{l}=0.11391$~au, $t_{p} \simeq 1.33$~fs): 
 (a) power spectrum $A_{z}(\omega)$ generated by the optically active $z$ degree of freedom; 
 (b) power spectrum $A_{\rho}(\omega)$ generated by the optically passive, transversal $\rho$ degree of freedom.} 
\end{figure} 
 
 Finally, in Figure~15, power spectra $A_{z}(\omega)$ and $A_{\rho}(\omega)$ are presented for the case when ${\rm H}_{2}^{+}$ 
 is excited by one-cycle laser pulse at the laser wavelength of $\lambda_{l}=400$~nm. 
 In the usual nomenclature of the harmonic order, three sequential peaks appeared in the power spectrum $A_{z}(\omega)$, 
 Figure~15(a), correspond respectively to the second, the fourth and the fifth harmonics with respect to the laser wavelength 
 of $\lambda_{l}=400$~nm used to excite ${\rm H}_{2}^{+}$. In a suggested new nomenclature, the second and the third peaks 
 (those at 98 and 82~nm) approximately correspond to the second harmonic of the smallest-frequency peak at 200~nm. 
 
 In the power spectrum $A_{\rho}(\omega)$, Figure~15(b), the peak at 190~nm corresponds to the second and that at 110~nm  
 approximately corresponds to the fourth harmonic with respect to the laser wavelength of $\lambda_{l}=400$~nm used. 
 On the other hand, the peak at 110~nm approximately corresponds to the second harmonic of the smallest-frequency peak  
 at 190~nm. Again, the peak at 190~nm corresponds to the frequency of $\omega^{\rho}_{1} \approx \omega_{\rm osc}$,  
 while the peak at 110~nm corresponds to the doubled frequency of $\rho$-oscillations  
 $\omega^{\rho}_{2} \approx 2\omega_{\rm osc}$. 
 
 \section{Conclusion} 
 
 In the present work, the non-Born-Oppenheimer quantum dynamics of ${\rm H}_{2}^{+}$ excited by linearly polarized along  
 the molecular axis shaped one-cycle laser pulses has been numerically studied at different laser-carrier frequencies 
 corresponding to the laser wavelengths of $\lambda_{l}=$25, 50, 100, 200, 300 and 400~nm. 
 The amplitudes of the one-cycles laser pulses have been optimized such that the energy of ${\rm H}_{2}^{+}$ at the end 
 of each pulse, $\langle E \rangle \simeq -0.515$~au, was close from the below to its dissociation threshold. 
 For the sake of completeness, some results were obtained by making use of two-cycle laser pulses as well. 
 The present work provides a detailed and in-depth extension of our previous study 
 \cite{ParKuehnBandr:jpcA.2016.H2p.post-field} 
 where shaped two-cycle laser pulses at the laser wavelengths of $\lambda_{l}=$800 and 200~nm were used  
 to excite ${\rm H}_{2}^{+}$ close to its dissociation threshold. 
 
 The basic results obtained in the present work are summarized as follows. 
 
 (I) The most efficient excitation of ${\rm H}_{2}^{+}$ by one-cycle shaped laser pulses, 
 which requires the minimum laser electric-field amplitude to achieve the required energy of about -0.515~au 
 at the end of the pulse, takes place at the laser wavelength of $\lambda_{l}^{\rm opt} \simeq 200$~nm corresponding 
 to the laser-carrier frequency of $\omega_{l}^{\rm opt} \simeq 0.2278$~au.  
 This frequency plays the role of a characteristic oscillation frequency $\omega_{\rm osc} \simeq 0.2278$~au  
 and manifests itself as the carrier frequency of temporally shaped post-laser-field oscillations of the 
 time-dependent expectation values $\langle z \rangle$ and $\langle \partial V/\partial z \rangle$ corresponding 
 to the optically active $z$ degree of freedom, which exist on a long timescale of at least 50~fs  
 after the ends of the pulses used to excite ${\rm H}_{2}^{+}$ initially. 
 The corresponding values for ${\rm H}_{2}^{+}$ excited by two-cycle shaped laser pulses 
 \cite{ParKuehnBandr:jpcA.2016.H2p.post-field} are almost identical.
 
 (II) The optically passive, transversal $\rho$ degree of freedom, which is excited only due to the wave properties 
 of the electron, also demonstrates post-laser-field oscillations of the time-dependent expectation values  
 $\langle \rho \rangle$ and $\langle \partial V/\partial \rho \rangle$ which occur at two basic frequencies 
 $\omega^{\rho}_{1} \approx \omega_{\rm osc}$ and $\omega^{\rho}_{2} \approx 2\omega_{\rm osc}$. 
 The latter frequency corresponds to the frequency-doubling of electronic $\rho$-oscillations as compared 
 to electronic $z$-oscillations described previously  
 \cite{Paramon:05cpl.HH-HD,%
 Paramon:07cp.HH-HD-Muon,%
 ParKuehnBandr:jpcA.2016.H2p.post-field}. 
 
 (III) A characteristic feature of power spectra generated by $z$ and $\rho$ degrees of freedom when ${\rm H}_{2}^{+}$ 
 is excited by one-cycle laser pulses is a small number of generated harmonics: not more than three relatively strong 
 harmonics can be observed in the power spectra presented in Section~3.3.  
 A similar feature was observed in the case when ${\rm H}_{2}^{+}$ was excited by two-cycle laser pulses 
 \cite{ParKuehnBandr:jpcA.2016.H2p.post-field}. 
 In a more general case, the number of generated harmonics correlates with the number of optical cycles used to excite 
 a molecule, with the overall trend being ``the smaller optical cycles used to excite a molecule is, the smaller number 
 of harmonics is generated''. Such a correlation was discussed in more detail in our previous work 
 \cite{ParKuehnBandr:jpcA.2016.H2p.post-field} 
 for both ordinary (electronic) and muonic molecules. 
 
 (IV) The positions of many peaks in power spectra presented in Section~3.3 are not equal to integer multiples of the laser 
 carrier frequency $\omega_{l}$, especially in the case of lower-order harmonic generation at  
 $\lambda_{l} < \lambda_{l}^{\rm opt} \simeq 200$~nm described in Section~3.3.1.  
 It is known 
 \cite{Gross:prl.2001.HHG} 
 that such harmonics may be be generated by single atoms and molecules due to resonance effects. 
 In the current case of one-cycle laser pulses used to excite ${\rm H}_{2}^{+}$ close to its dissociation threshold 
 and the resonant properties of this process depicted in Figure~2, the situation can be rationalized by introducing 
 the suggested above new nomenclature of the harmonic order $\omega/\omega_{l}$ as follows. 
 Since the smallest-frequency harmonics in all power spectra $A_{z}(\omega)$ and $A_{\rho}(\omega)$ presented in Section~3.3  
 correspond to the optimal wavelength $\lambda_{l}^{\rm opt} \simeq 200$~nm, a new harmonic order $(\omega/\omega_{l})'$ 
 can be defined as follows: 
 \begin{equation} 
 \left(\frac{\omega}{\omega_{l}}\right)' =  
 \left(\frac{\omega}{\omega_{l}}\right) \frac{\lambda_{l}^{\rm opt}}{\lambda_{l}}. 
 \label{new-harm-order}  
 \end{equation} 
 
 It is easy to check that Equation~(\ref{new-harm-order}) works very well for all power spectra $A_{z}(\omega)$ and is  
 also quite suitable for power spectra $A_{\rho}(\omega)$ where the frequencies of strongest harmonics are given by 
 $\omega^{\rho}_{1} \approx \omega_{\rm osc}$ and $\omega^{\rho}_{2} \approx 2\omega_{\rm osc}$. 
 
 \section*{Acknowledgement}
  This work has been financially supported by the Deutsche Forschungsgemeinschaft
  through the Sfb 652 (O.K.), which is gratefully acknowledged.

\end{document}